\documentstyle[aps,twocolumn,psfig]{revtex}

\def\bfv{{\bf v}}
\def\de{{\delta}}
\def\ep{{\epsilon}}
\def\ga{{\gamma}}
\def\om{{\omega}}
\def\re#1{(\ref{#1})}
\newcommand{\bfna}{{\mbox{\boldmath $\bf\nabla$}}}
\newcommand{\sign}{{\rm sign}}                      
\newcommand{\wh}[1]{{\widehat{#1}}}
\newcommand{\wt}[1]{{\widetilde{#1}}}

	\def\figi{
	\begin{figure}
	\def\wi{1.7in}
	\centerline{\psfig{file=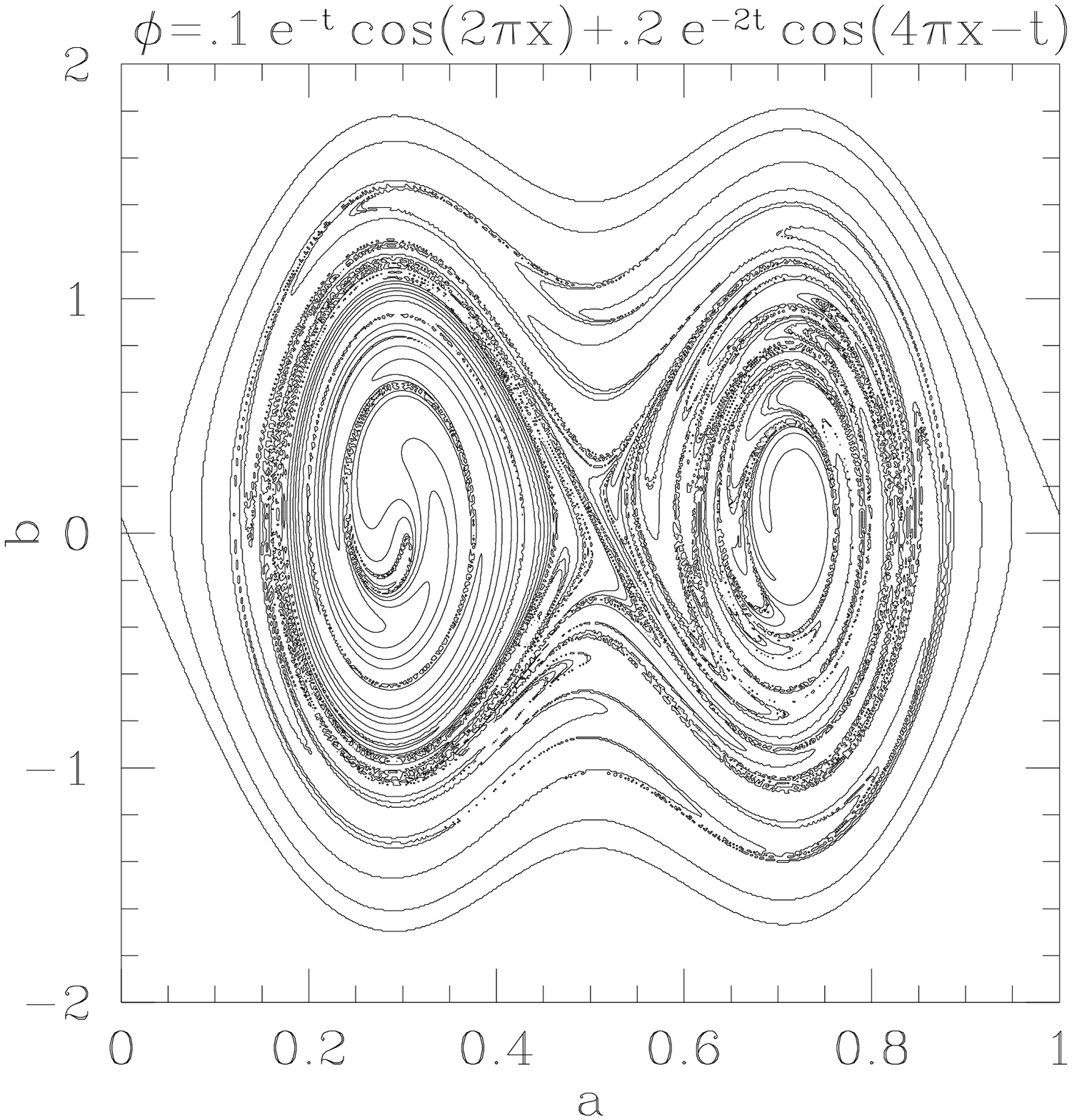,width=\wi}\psfig{file=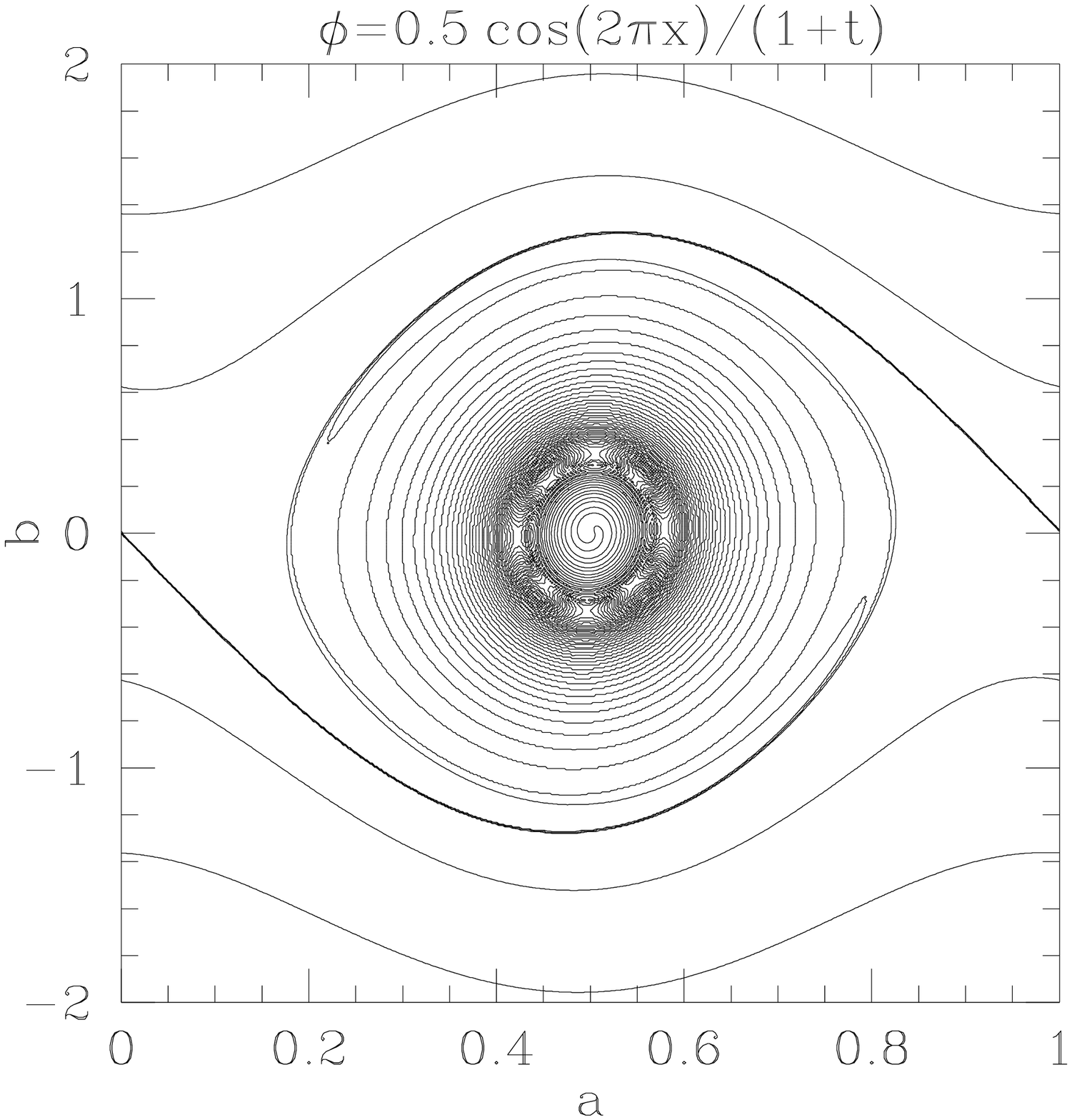,width=\wi}}
	\centerline{\psfig{file=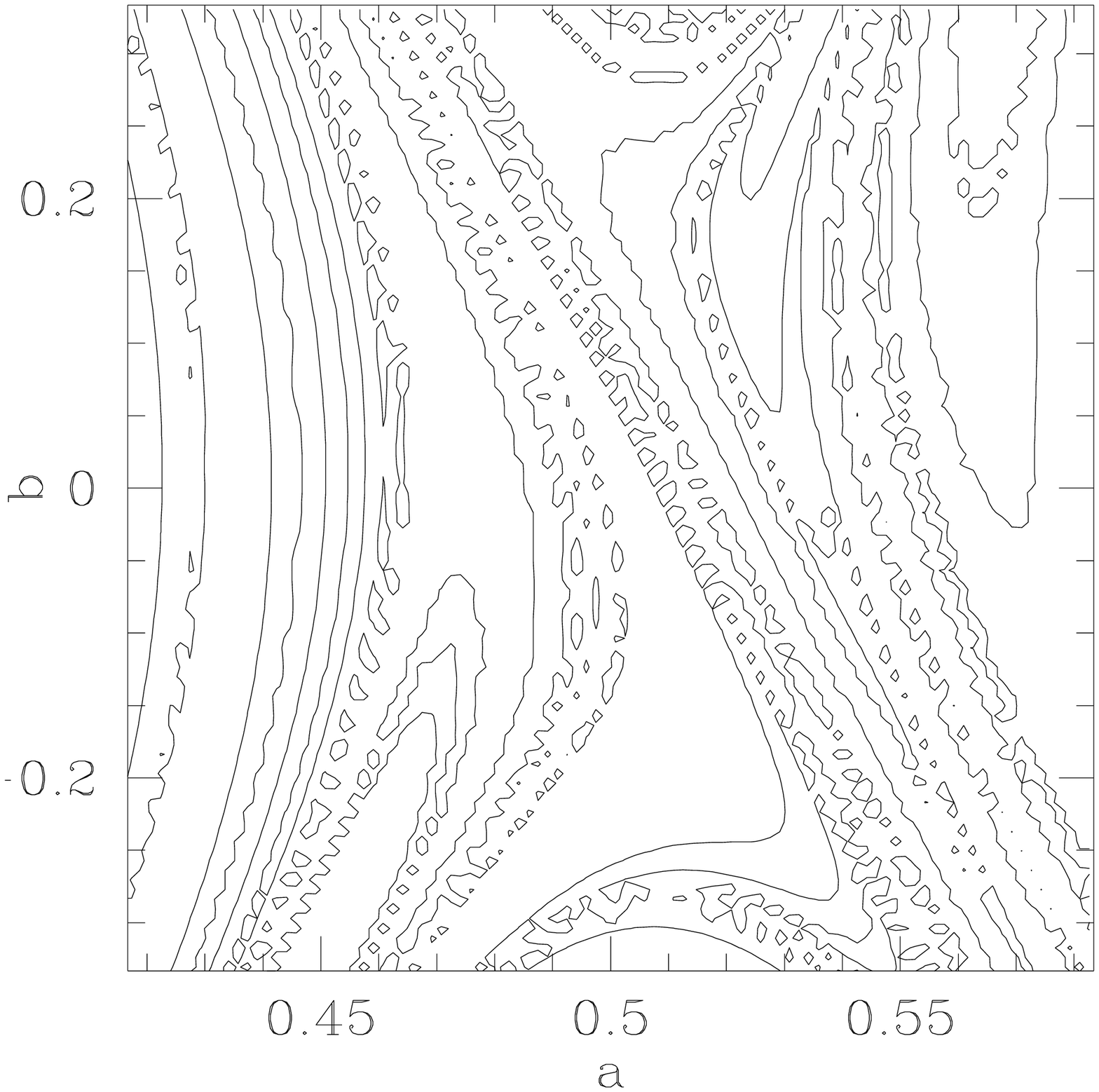,width=\wi}\psfig{file=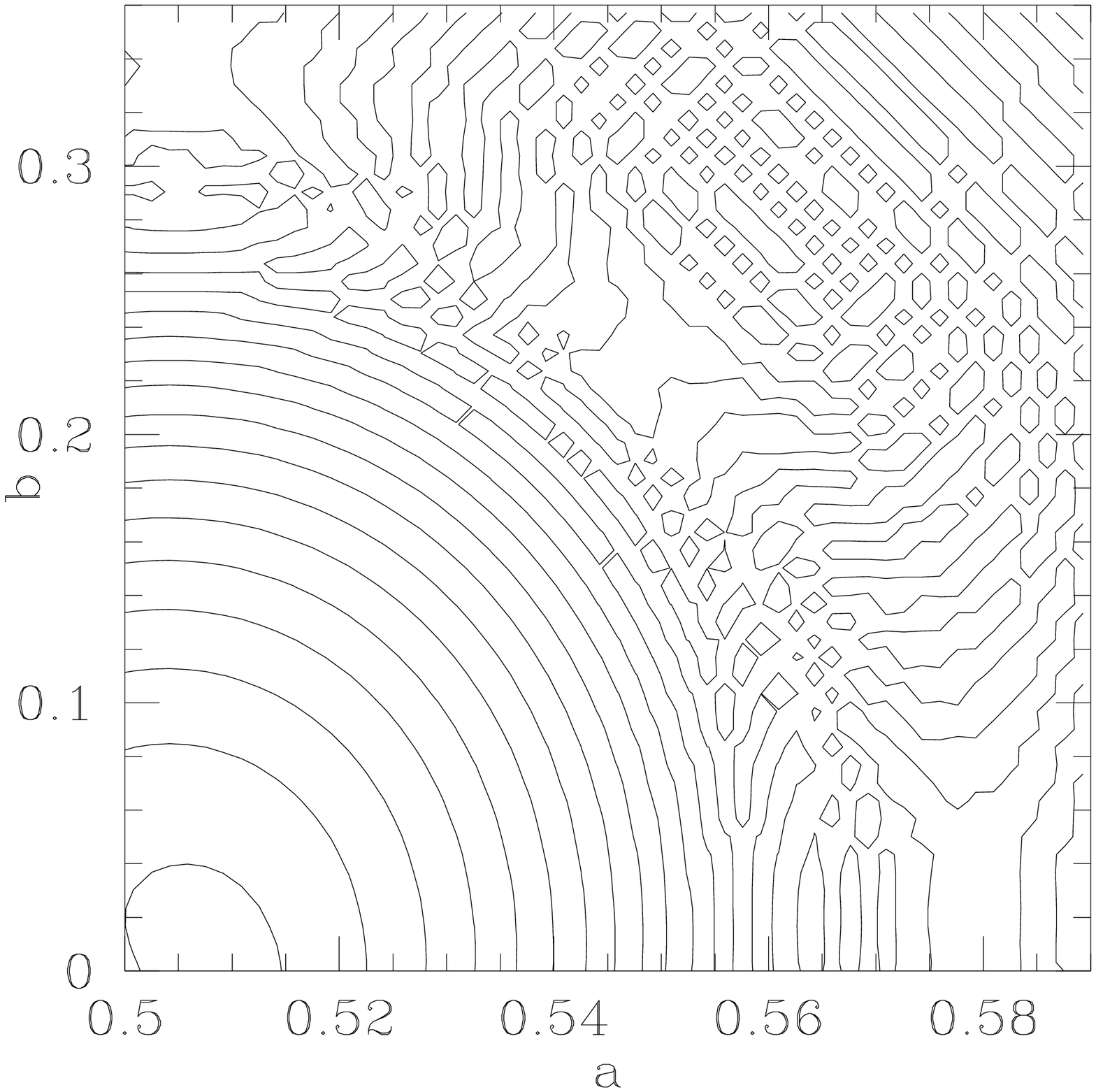,width=\wi}}
	\begin{small}
	\caption{Contour lines of $U(a,b)$ computed for exponentially (left)
	and algebraically (right) decaying potentials.  The computation was
	done through the ``infinite'' time $t=20$ and $t=1000$, respectively.
	The bottom is a zoom of the top.  The presense of multiple extrema
	(O-points) and saddles (X-points) of $U$ is apparent.}
	\label{fig:1}
	\end{small}
	\end{figure}
	}

	\def\figii{
	\begin{figure}
	\def\wi{3in}
	\centerline{\psfig{file=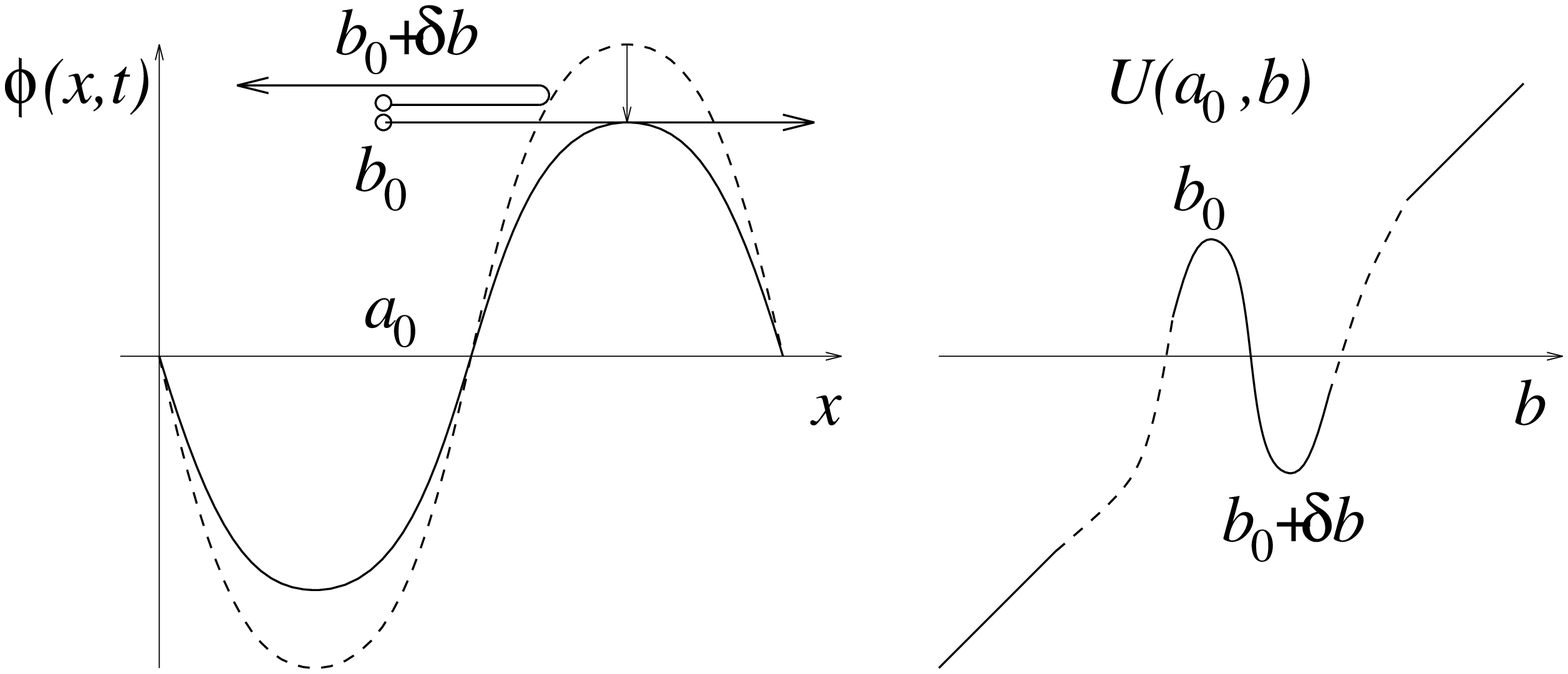,width=\wi}}
	\vspace{0.1in}
	\begin{small}
	\caption{Particle bouncing in a decaying potential and its signature
	$U(a,b)$.  Near the potential top, an increase in the initial velocity
	$b$ can bring the particle to the decaying potential barrier earlier,
	when the barrier was higher, and thus turn the particle around:
	$U(a_0,b_0)>0$, $U(a_0,b_0+\de b)<0$.}
	\label{fig:2}
	\end{small}
	\end{figure}
	}

	\def\figiii{
	\begin{figure}
	\def\wi{2.5in}
	\centerline{\psfig{file=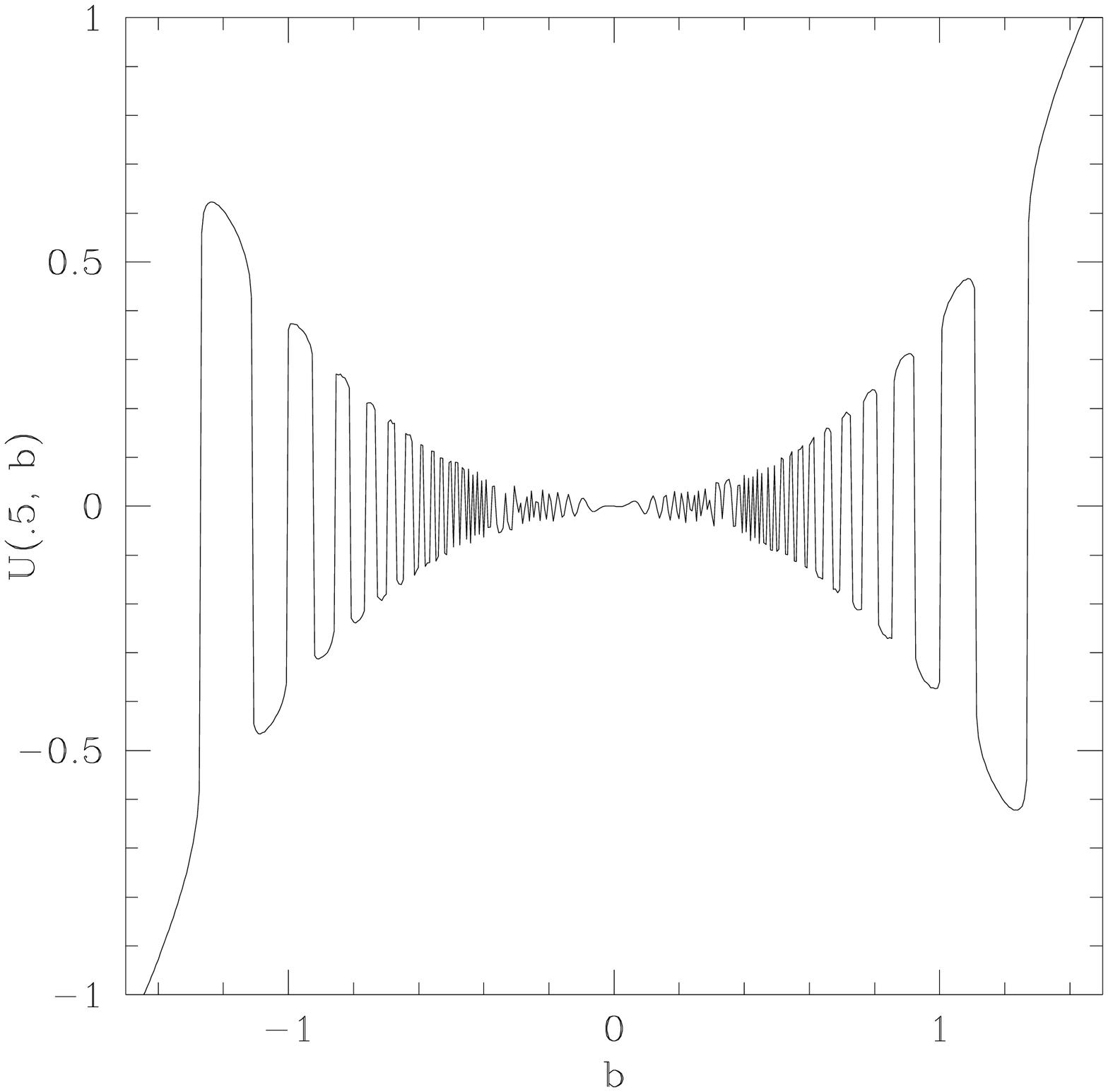,width=\wi}}
	\begin{small}
	\caption{The cross-section of $U(a,b)$ for the algebraically decaying
	potential shown in Fig.~\ref{fig:1}.  Near the bottom of the
	potential well,	the particle makes many bounces before being
	released in an essentially random direction.}
	\label{fig:3}
	\end{small}
	\end{figure}
	}

	\def\figiv{
	\begin{figure}
	\def\wi{3.2in}
	\centerline{\psfig{file=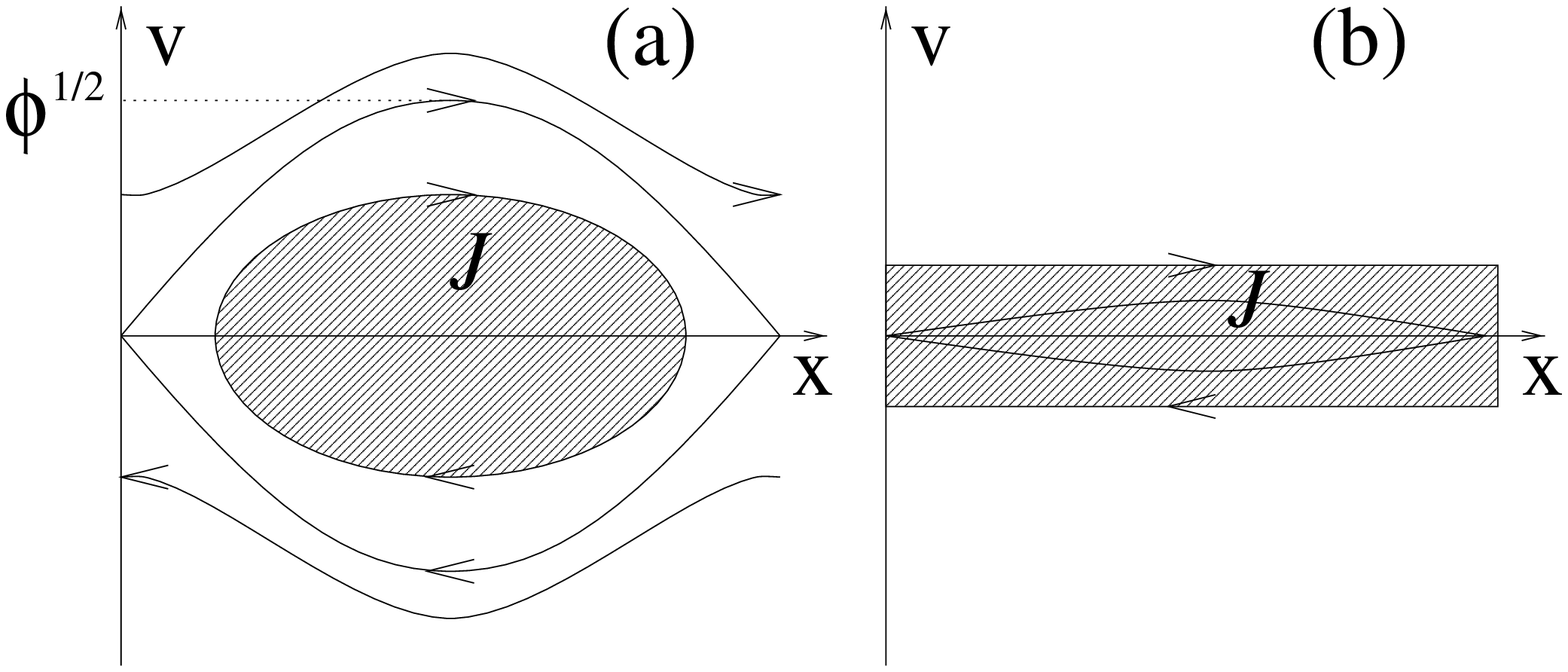,width=\wi}}
	\begin{small}
	\caption{The adiabatic invariant before (a) and after (b) the
	separatrix crossing in a decaying potential well.}
	\label{fig:4}
	\end{small}
	\end{figure}
	}

\begin{document}
\bibliographystyle{../macros/prsty}

\title{Nonlinear Landau damping in collisionless plasma and inviscid
fluid} 
\author{M.~B. Isichenko\\
{\small\it Physics Department, University of California--San Diego,
La Jolla, CA 92093-0319}} 
\date{\small December 12, 1996} 
\maketitle

\begin{abstract}
{\em Abstract.}  The evolution of an initial perturbation in Vlasov
plasma is studied in the intrinsically nonlinear long-time limit
dominated by the effects of particle trapping.  After the possible
transient linear exponential Landau damping, the evolution enters into
a universal regime with an algebraically damped electric field,
$E\propto1/t$.  The trick used for the Vlasov equation is also applied
to the two-dimensional (2D) Euler equation.  It is shown that the
stream function perturbation to a stable shear flow decays as
$t^{-5/2}$ in the long-time limit.  These results imply a strong
non-ergodicity of the fluid element motion, which invalidates
Gibbs-ensemble-based statistical theories of Vlasov and 2D fluid
turbulence.

PACS 52.35.Ra, 47.27.-i, 47.15.Ki.
\end{abstract}

Part of the challenge facing the theory of turbulence is that it
is extremely difficult to make exact statements about the long-time
behavior of a nonintegrable system that go beyond the mere
consequences of applicable conservation laws. For chaotic systems with
a few degrees of freedom, there are a few results like this, including
the little-known Sundman's theorems for the three-body problem
cf.~\cite[pp.~49--68]{SM71} and the famous Kolmogorov-Arnold-Moser
theory \cite{Arnold63-transl}.  Here an attempt is made to draw
certain long-time conclusions about the nonlinear evolution in a
Vlasov plasma and in a 2D ideal fluid.  We study the dynamics of the
relaxation of a generic initial perturbation in these systems and
derive algebraic damping laws for the perturbation.  As in the above
finite-dimensional examples, our continuous findings imply the lack of
ergodicity, with grave implications for several statistical theories
of turbulence.

We start with the Vlasov-Poisson system for the electron distribution
function $f(x,v,t)=f_0(v)+\wt f(x,v,t)$ and the electric field
$E(x,t)=-\partial_x\phi(x,t)$,
\begin{equation}
(\partial_t+v\partial_x+E\partial_v)f=0, \quad
\partial_xE=\int_{-\infty}^\infty f\,dv-1,
\label{vlasov}
\end{equation}
describing nonlinear plasma waves on a uniform ion background.  In
Eq.~\re{vlasov}, the time $t$ is normalized to the inverse plasma
frequency $\om_{pe}^{-1}$, and $x$ is measured in Debye lengths
$r_D=v_e/\om_{pe}$, where $v_e$ is the electron thermal velocity, the
unit for $v$.  The problem has two basic dimensionless parameters, the
nonlinearity $\ep\sim\wt f/f_0$ and the wavenumber $kr_D$ of the
initial perturbation.

The original solution of the initial-value problem for Vlasov plasma
by Landau \cite{Landau46} is strictly linear, meaning that $\ep$ is
the smallest parameter of the problem.  We will not assume either of
the parameters $\ep$ or $k$ small or large; instead, the largest, or
the only large, parameter in our treatment will be time.  The
long-time limit is intrinsically nonlinear, because the linearization
of the Vlasov-Poisson system fails for $t$ larger than the particle
bounce time $\tau_b\simeq\ep^{-1/2}$ \cite{Oneil65}.  This happens
because the fluctuations of the distribution function do not decay,
but rather develop free-streaming-type small scales, $\wt
f(x,v,t)\simeq\wt f(x-vt,v,0)\sim\ep$, and the nonlinearity, $\partial_v\wt
f/f'_0(v)\sim\ep t$, increases secularly with time.

The previous analytical work on the nonlinear Vlasov plasma includes
the exact special stationary solutions of Bernstein, Greene, and
Kruskal (BGK) \cite{BGK57} and the non-stationary theory of O'Neil for
$\tau_b\le t\ll\ep^{-1}$ and $k\ll1$ \cite{Oneil65}.  O'Neil showed
that, due to trapping and phase mixing, the damping rate of the wave,
$\ga(t)=\dot\phi/\phi$, starts oscillating about zero with the time
scale $\tau_b$ and a decreasing amplitude.  The currently prevailing
conjecture is that nonlinear plasma waves, after several such
oscillations, settle to a stationary stable BGK wave.  This conclusion
appears to be backed by numerical simulation \cite{GIBFFS88}, although
numerical evidence should not be considered conclusive for the
long-time limit. More importantly, the stability of the nonlinear BGK
waves remains an outstanding issue.  This author is not aware of any
single example of a stable BGK wave; moreover, all analytically
written BGK waves appear linearly unstable \cite{GIBFFS88}, and the
only known nonlinear stability criterion \cite{Hohl69}, $df_0/dH<0$,
where $H(x,v)=\phi(x)+v^2/2$ is the particle energy, holds for no
periodic BGK wave \cite[P.~85]{Davidson72}.  This suggests that the
Landau damping will not be arrested by nonlinearity; however, the
nature of the damping will be modified for large $t>\ep^{-1}$.

Our logic is as follows.  We {\em assume\/} that the electric field
decays with time: $E(x,t)\to0$, as $t\to\infty$.  Then this assumption
is shown to be self-consistent by calculating the actual damping rate,
$E\propto t^{-1}$, instead of the linear exponential damping.

Assume a periodic boundary condition in $x$ with the period $L$, and
expand the electric field $E$ in a Fourier series.  Then, for $k\ne0$,
the second Eq.~\re{vlasov} yields:
\begin{eqnarray}
&&ikE_k(t)=(2\pi L)^{-1}\int_{-\infty}^\infty
dv\int_0^Ldx\,f(x,v,t)e^{-ikx}
\nonumber\\
&&=(2\pi L)^{-1}\int f_i(a,b)e^{-ikx(a,b,t)}\,da\,db.
\label{E1}
\end{eqnarray}
In Eq.~\re{E1}, the variables of integration were changed to the
Lagrangian variables $a$ and $b$, the initial position and the
velocity of a particle.  According to the Liouville theorem, the
Jacobian of this transformation is unity, and the distribution
function is constant along the particle orbit thus reducing to its
initial value $f_i(a,b)\equiv f(a,b,0)$.

Equation \re{E1} expresses the electric field in terms of the particle
orbit $x(a,b,t)$ defined by $\ddot x=E(x,t)$ and the given initial
condition, a problem as difficult as the original Eq.~\re{vlasov}.
However, the integral representation of $E$ in terms of the orbit is
very useful for studying the long-time asymptotic, when the electric
field is presumably small, and the orbit becomes a motion with a
constant velocity, $x(a,b,t)=U(a,b)\,t$ (plus lower-order terms).  The
resulting integral of an oscillatory function,
\begin{equation}
E_k(t)\propto\int f_i(a,b)e^{-ikt\,U(a,b)}\,da\,db,
\quad t\to\infty,
\label{E2}
\end{equation}
for smooth $f_i$, will generally have only two kinds of asymptotics.
If the gradient of $U(a,b)$ is nowhere zero (as, for example, in the
linear theory, where $U\simeq b$), then the integral \re{E2} is
exponentially small at large $t$ (the Riemann-Lebesgue lemma).  If, on
the other hand, $U$ has a stationary point where $\partial_aU=\partial_bU=0$, then
the $O(t^{-1/2})$-vicinity of this point dominates the integral, which
scales as $E\propto t^{-1}$.  Below we show that $U(a,b)$ has
stationary points in the general case, and therefore $E$ decays
algebraically.

\figi
The problem of finding the final velocity $U$ as a function of the
initial condition, for a particle moving in a decaying potential, is
very similar to chaotic scattering \cite{Eckhardt88}, and, likewise,
due to the transient particle trapping, the function $U(a,b)$ is quite
complex (Fig.~\ref{fig:1}).  We are interested in whether $U(a,b)$ is
a monotonic function of its arguments.  The fact that it is not is
most transparent from the inspection of the particle bouncing at the
top (Fig.~\ref{fig:2}) and at the bottom (Fig.~\ref{fig:3}) of a
decaying potential profile.  If the initial potential amplitude is
small, the bouncing at the bottom is possible only if $\phi$ decays
sufficiently slowly, e.g., $\phi\propto \ep\,t^{-\alpha}$,
$0<\alpha<2$, in order that the bounce time $\tau_b\propto\phi^{-1/2}$
be less than $t$. The initial, linear Landau damping is exponential,
seemingly suggesting no bouncing, hence no stationary points of
$U(a,b)$ and the persistence of the exponential damping.  However, a
simple perturbation analysis of the particle motion {\em near the
top\/} of an evolving potential hill shows that one can always pick
initial conditions such that the behavior of Fig.~\ref{fig:2} takes
place.  To some confusion, this turns out possible only if the spatial
extrema of $\phi(x,t)$ and $\partial_t\phi(x,t)$ do not coincide; that
is, if there is more than just one wave, a safely generic situation.
(The result of the left Fig.~\ref{fig:1} is for two potential waves.
A similar computation for one wave shows a smooth $U$ with no
stationary points.)  \figii \figiii

In fact, $U(a,b)$ has an infinite number of stationary points
$(a^j,b^j)$.  Upon expanding the particle orbit near such a point at
large $t$, $x(a,b,t)=U^jt+\left[U^j_{aa}(a-a^j)^2+U^j_{bb}(b-b^j)^2+
U^j_{ab}(a-a^j)(b-b^j)\right]t/2+V^j\ln t+W^j+O(t^{-1})$, Eq.~\re{E1}
yields the electric field at large $t$ in terms of the infinite
series,
\begin{equation}  
E_k=\sum_j
\frac{f_i(a^j,b^j)\,e^{-ik(U^jt+V^j\ln t+W^j)}}
{k^2Lt\,\left[U^j_{aa}U^j_{bb}-(U^j_{ab})^2\right]^{1/2}}+
O\left(\frac{1}{kt^2}\right),
\label{Eseries}
\end{equation}
which could in principle pose problems in terms of divergencies or
cancellations.  
\figiv

The series \re{Eseries} turns out to be absolutely (exponentially in
$j$) convergent, because it is possible to analyze the accumulation of
the stationary points of $U$.  This is due to the adiabaticity of the
particle motion at large time, when the bounce frequency
$\om_b\propto\phi^{1/2}\propto t^{-1/2}$ is much larger than the
potential damping rate $\dot\phi/\phi\propto t^{-1}$.  As a result, the
adiabatic invariant $J(a,b)$, the $(x,v)$-plane area inside a nearly
closed trapped particle orbit, is conserved, and the corresponding
angle variable $\theta$ is growing with the bounce frequency:
$\theta=\int^t\om_bdt\propto t^{1/2}$.  Untrapping occurs when the
shrinking separatrix of the decaying potential, with the area
$S\propto t^{-1/2}$, intersects the orbit with the conserved area $J$
(Fig.~\ref{fig:4}).  For a small $J\propto (a-a_0)^2+b^2$, the
crossing time $t^*\propto J^{-2}$ and the angle $\theta^*\propto J^{-1}$.
Following a small change during the separatrix crossing \cite{TCE86},
the adiabatic invariant of the passing particle (now defined as twice
the phase-space area) is conserved again and defines the final
velocity $|U(a,b)|\simeq J(a,b)/(2L)$.  The sign of $U$, roughly
$\sign(\sin\theta^*)$, depends on whether the crossing happens in the
upper or in the lower half-plane of Fig.~\ref{fig:4}.  The width of
the steps of $U$ is still finite, $\de\theta^*\propto
e^{-\om_bt^*}\propto e^{-1/J(a,b)}$; it is determined by the
exponentially narrow near-separatrix layer, where the bounce period
$2\pi/\om_b$ diverges logarithmically, and the adiabaticity does not
hold.  Thus we obtain the approximate analytical expression for the
final velocity:
\begin{equation}
U(a,b)\simeq J(a,b)/(2L)\,\tanh\left[e^{1/J(a,b)}\sin J^{-1}(a,b)\right].
\label{U}
\end{equation}
Near the bottom of the well, $a=a_0$, the behavior of Eq.~\re{U} is
consistent with the numerical result in Fig.~\ref{fig:3}.  Equation
\re{U} also implies the exponentially growing curvature
$U^j_{bb}\propto e^j$ near the steps as one moves to the accumulation
point of the $U$ extrema, hence the exponential convergence of the
series \re{Eseries}.

Thus the long-time behavior of the electric field \re{Eseries} is
dominated by a few ``strongest'' stationary points of $U(a,b)$.  In
addition to the algebraic damping rate, we infer as a by-product the
spectrum $E_k\propto k^{-2}$, $k\ll t$, implying the development of
steps in the electron density perturbation $\partial_xE$.

We now turn to the different problem of the relaxation in 2D ideal
inviscid incompressible fluid with the velocity
$\bfv=\bfna\psi(x,y,t)\times\wh{\bf z}$ described by the Euler equation,
\begin{equation}
(\partial_t+\bfv\cdot\bfna)\,\om=0,
\quad
\om=-\bfna^2\psi.
\label{euler}
\end{equation}
As in the case of Vlasov plasma, we are interested in the long-time
relaxation of an initial perturbation $\wt{\psi}(x,y,t)$ imposed on a
stable shear flow $\psi_0(x)$.  The deep analogy of this problem with
the Landau damping in plasmas has been noted \cite{BDL70,BdC-NY96}.
We will assume a periodic boundary condition in $y$ [along the shear
flow $v_0(x)=-\psi_0'(x)$].  In linear theory, the perturbation of the
stream function $\psi$ is known to be damped, because the
reconstruction of $\psi$ from the conserved vorticity $\om$ with
growing gradients involves an integration,
\begin{equation}
\psi(x,y,t)=
\int G(x,x',y-y')\,\om(x',y',t)\,dx'\,dy',
\label{ps1}
\end{equation}
where $G$ is the boundary-condition-dependent Green's function with a
discontinuous derivative at $(x,y)=(x',y')$.  If the flow is unbounded
in the $x$ direction, for example, $G_k(x,x')=e^{-|k(x-x')|}$.  Unlike
plasma waves, the damping law of $\wt{\psi}$ is algebraic already in
linear theory: $\wt{\psi}\propto t^{-2}$ for monotonic $v_0(x)$
\cite{Yamagata76,Brown-Stewartson-80,BdC-NY96} and $\propto t^{-1/2}$
for $v_0(x)$ with an extremum \cite{Brunet-Warn-90}.

Similarly to the Vlasov case, linear approximation in the Euler
equation breaks for large $t$ raising the question of the long-time
asymptotic.  To this end, we use the same trick as for the Vlasov
equation.  Upon applying the Fourier transform in $y$ to Eq.~\re{ps1}
and changing the integration variables $(x',y')$ to the Lagrangian
variables $(a,b)$, we obtain:
\begin{equation}
\psi_k=
\int G_k[x,X(a,b,t)]\,e^{-ikY(a,b,t)}\,\om_i(a,b)\,da\,db,
\label{ps2}
\end{equation}
where $\om_i(a,b)$ is the total initial vorticity, and $(X,Y)$ is the
orbit of a fluid element with the initial position $(a,b)$.  Consider
the case of a smooth and monotonic $v_0(x)$.  Then, for a very small
perturbation, the unperturbed orbit $(X,Y)\simeq(a,b+v_0(a)\,t)$
yields an oscillatory integral in $a$, which is not exponentially
small because of the derivative discontinuity in $G_k$.  Changing the
variable $a$ to the monotonic $v_0(a)$ and integrating by parts twice
then yields $\psi_k(x,t)\propto t^{-2}$ for $k\ne0$, the well-known
linear result.  For $v_0(x)$ with a stationary point, the singularity
of the Green's function does not matter, and a stationary-phase
integration over $a$ yields $\wt{\psi}\propto t^{-1/2}$, in agreement
with \cite{Brunet-Warn-90}.  Based on the ordering of terms in
Eq.~\re{euler} for the regime with $\wt{\psi}\propto t^{-2}$, Brunet and
Warn \cite{Brunet-Warn-90} argued that the nonlinearity remains small
and does not change the damping rate.  Such an analysis appears
superficial, because the accumulation of small nonlinear effect in the
Euler equation is secular.  Our analysis of Eq.~\re{ps2} goes as
follows.  The flow disturbance of order $\ep$ makes the orbit
essentially depend on both $a$ and $b$, e.g., $Y=b+v_0(a)\,t+\ep \int
v_1(a,b,t,\ep)\,dt$.  Thus the integral \re{ps2} is also oscillatory
in $b$ for $t>\ep^{-1}$.  This is when the nonlinearity comes into
effect.  Because of periodicity, the phase $iktY(a,b)$ has a
stationary point in $b$ producing the additional factor of $t^{-1/2}$
in the integral asymptotic.  Finally, for a smooth stable monotonic
shear velocity profile, a smooth stream function perturbation decays
as $\wt{\psi}\propto t^{-5/2}$ for $t\gg\ep^{-1}$.  Similarly, for
$v_0(x)$ with an extremum, $\wt{\psi}\propto t^{-1}$.

One of the interesting consequences of the nonlinear Landau damping
concerns ergodicity, the assumption underlying the Gibbs-ensemble
theories of turbulence in the Vlasov-Poisson system
\cite{Lynden-Bell67,THL-B86} and in 2D fluid
\cite{Miller90,RS91,MWC92}.  In these theories, the statistical
ensemble includes all possible permutations of phase-space (fluid)
elements with the associated distribution function $f$ (vorticity
$\om$), via either combinatorial treatment or path integration for the
partition function.  Such analyses predict specific quantitative
results, such as the final relaxed state, for an arbitrary initial
condition.  In addition to the difficulties with non-Gaussian path
integrals \cite{Isichenko-Gruzinov-94}, the Gibbs-ensemble
theory cannot be true in such a generality because of the nonlinear
Landau damping.  For example, if the initial condition is a slightly
perturbed stable shear flow $v_y(x)$, the zonal velocity $v_x(x,y,t)$
will decay as $\ep t^{-5/2}$, and the zonal displacement of any fluid
element, $\de X=\int v_xdt$, is for ever bounded by a small constant:
$|\de X|<C\ep$.  This purely dynamical fact does not follow from
conservation laws alone and implies that the fluid element motion is
strongly non-ergodic.  It follows that the existing statistical
theories do not work, at least for initial conditions close to stable
shear flows. The same is true of the Vlasov-Poisson system, where the
velocity change for any particle, $\de v=\int E(x(t),t)\,dt$, is
bounded for infinite time, because $E\propto t^{-1}$, and the
convergence of $\de v$ is ensured by another power of $t$ coming from
the nearly uniform motion in the coordinate $x\simeq Ut$, over which
$E$ is zero-average.  Again, heating up a small group of particles to
arbitrarily high energies in an evolving collisionless plasma does not
contradict conservation laws; however, this turns out exactly
prohibited by self-consistent dynamics.

The shear damping of perturbations is not specific to plane parallel
flows in the Euler equation; quite similar results must hold for
circular monopole vortices developing in the course of long-time
turbulent evolution \cite{CMcWPWY91,WMcW92} and also in the framework
of related 2D geophysical fluid equations, where the nonlinear Landau
damping is the mechanism of the turbulence relaxation toward
large-scale coherent structures.  Finally, it appears that the
decaying 2D turbulence is more about dynamics (vortex merger and the
nonlinear damping of vortex perturbations) than statistics.

{\em Acknowledgments.}  This work grew out of extensive and fruitful
discussions with Andrei Gruzinov, who suggested the possible
non-ergodicity of ideal fluid flows.  I also thank T.~M. O'Neil,
J.~M. Greene, and W.~R. Young for very useful and pleasant
conversations.  This work was supported by the U.S.\ DOE Grant
No.~DE-FG0388ER53275 and ONR Grant No.~N00014-91-J-1127.

\bibliography{../bib/turb,../bib/chaos,../bib/mbi,../bib/plasma}
\end{document}